\documentclass[useAMS,usenatbib]{mnras}
\usepackage{amsmath}
\usepackage{amssymb}
\usepackage{amsfonts}
\usepackage{graphicx}
\usepackage[dvipsnames]{xcolor}
\hypersetup{colorlinks=true,allcolors=teal}

\usepackage{setspace}
\usepackage{microtype}

\newcommand{\p}{\ensuremath{\partial}}

\newcommand{\Msun}{\ensuremath{M_{\odot}}}

\newcommand{\Mpc}{\ensuremath{\,{\rm Mpc}}}
\newcommand{\kpc}{\ensuremath{{\,\rm kpc}}}
\newcommand{\kms}{\ensuremath{{\rm \,km\,s}^{-1}}}

\newcommand{\e}[1]{\ensuremath{{\rm e}^{#1}}}

\newcommand{\eqn}[1]{equation~\eqref{#1}}

\newcommand{\be}{\begin{equation}}
\newcommand{\ee}{\end{equation}}

\newcommand{\bvec}[1]{\ensuremath{\boldsymbol{#1}}}

\title[Star formation and tides]
{
\emph{A sea of tides:} star formation and the central-satellite dichotomy in a continuum of tidal environments}

\author[Zjupa et al.]
{Jolanta Zjupa$^{1}$\thanks{E-mail: jolanta.zjupa@oca.eu},
 Aseem Paranjape$^{2}$,
 Oliver Hahn$^{1}$, 
 R\"{u}diger Pakmor$^{3}$
 \vspace*{0.2cm}\\
  $^1$ Laboratoire Lagrange, Universit\'e C\^ote d'Azur, Observatoire de la C\^ote d'Azur, CNRS, Blvd de l'Observatoire,\\\hskip0.15in CS 34229, 06304 Nice cedex 4, France\\
  $^2$ Inter-University Centre for Astronomy \& Astrophysics, Ganeshkhind, Post Bag 4, Pune 411007, India\\
  $^3$ Max-Planck-Institut f\"{u}r Astrophysik, Karl-Schwarzschild-Str. 1, D-85748, Garching, Germany
} 
 
\begin{document}

\date{draft}
\pubyear{2020}

\pagerange{\pageref{firstpage}--\pageref{lastpage}} 

\maketitle

\label{firstpage}

\begin{abstract}
The environment-dependent bimodality of the distribution of stellar mass ($M_\ast$) and specific star formation rate (sSFR) of galaxies, and its explanation in terms of the central-satellite dichotomy, form a cornerstone of our current understanding of galaxy evolution in the hierarchical structure formation paradigm. We revisit this framework in the IllustrisTNG simulation in the context of the \emph{most extreme} local tidal anisotropy $\alpha_{\rm peak}$ experienced by each galaxy over cosmic time, which is an excellent proxy for environmental influence. We show that, while sharing a common monotonic $M_\ast$-$v_{\rm peak}$ relation, central, satellite and `splashback' galaxies define a \emph{hierarchy of increasing} $\alpha_{\rm peak}$. We also find that the sSFR of objects in small haloes unaffected by feedback from an active nucleus typically decreases with increasing $\alpha_{\rm peak}$. Our results support an alternate viewpoint in which a galaxy can be identified by the value of $\alpha_{\rm peak}$; i.e., rather than being placed on the central-satellite dichotomy, a galaxy is better classified by its location in a continuum of tidal environments. This conceptual shift can potentially yield a more robust understanding of galaxy evolution and the galaxy-dark matter connection, e.g., in accurately modelling subtle effects such as sSFR-induced secondary clustering.
\end{abstract}

  
\section{Introduction} 
\label{sec:intro}
\noindent

The persistent bimodality of the galaxy population in the plane of stellar mass ($M_\ast$) and specific star formation rate (sSFR = SFR/$M_\ast$) and its dependence on the cosmic environment as revealed by large-volume surveys \citep[e.g.,][]{baldry+04,kauffmann+04,zehavi+11}, have played a fundamental role in shaping our current understanding of galaxy formation and evolution. Observed galaxies primarily segregate into relatively weakly clustered star forming objects and strongly clustered `quenched' objects. The origin of these trends, and their connection to the cosmic web, has been a subject of intense research \citep[e.g.,][]{dressler80,weinmann+06,peng+12,woo+17,poudel+17}.

Explanations of this bimodality typically invoke the \emph{central-satellite dichotomy}. In this framework \citep[see, e.g.,][and references therein]{ss09}, a `central' galaxy forms and resides close to the gravitational centre of an independent host dark matter halo \citep{wr78}, with its star formation regulated by feedback from supernovae or an active galactic nucleus (AGN) depending on its mass \citep{db06}. On the other hand, a `satellite' resides in a subhalo of a larger host, having been a central before the merger event that transformed its own host into a subhalo. The (typically crowded and hot) environment of the group host leads to additional satellite-specific processes such as strangulation, ram pressure stripping, and others, which effectively decrease the amount of fuel available for star formation \citep[see, e.g.,][and references therein]{vdb+08}. Star formation activity thus strongly correlates with galaxy type and hence environment: low-mass star forming objects are predominantly centrals and hence relatively weakly clustered compared to quenched objects of similar $M_\ast$ which are predominantly satellites of more massive (and typically AGN-quenched) centrals.

\begin{figure*} 
\centering
\includegraphics[width=0.45\textwidth,trim= 8 8 6 8,clip]{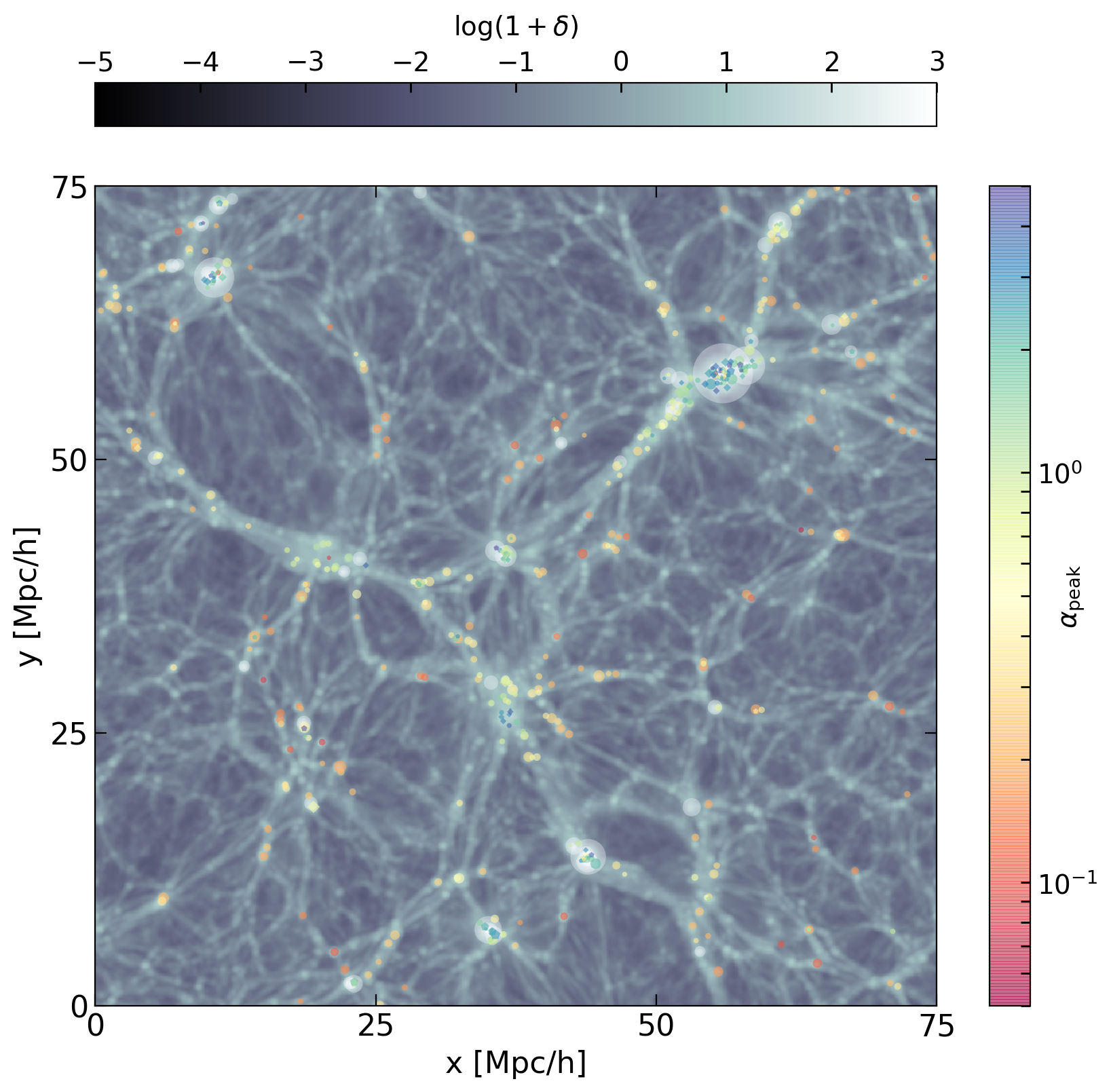}
\includegraphics[width=0.45\textwidth,trim= 8 8 6 8,clip]{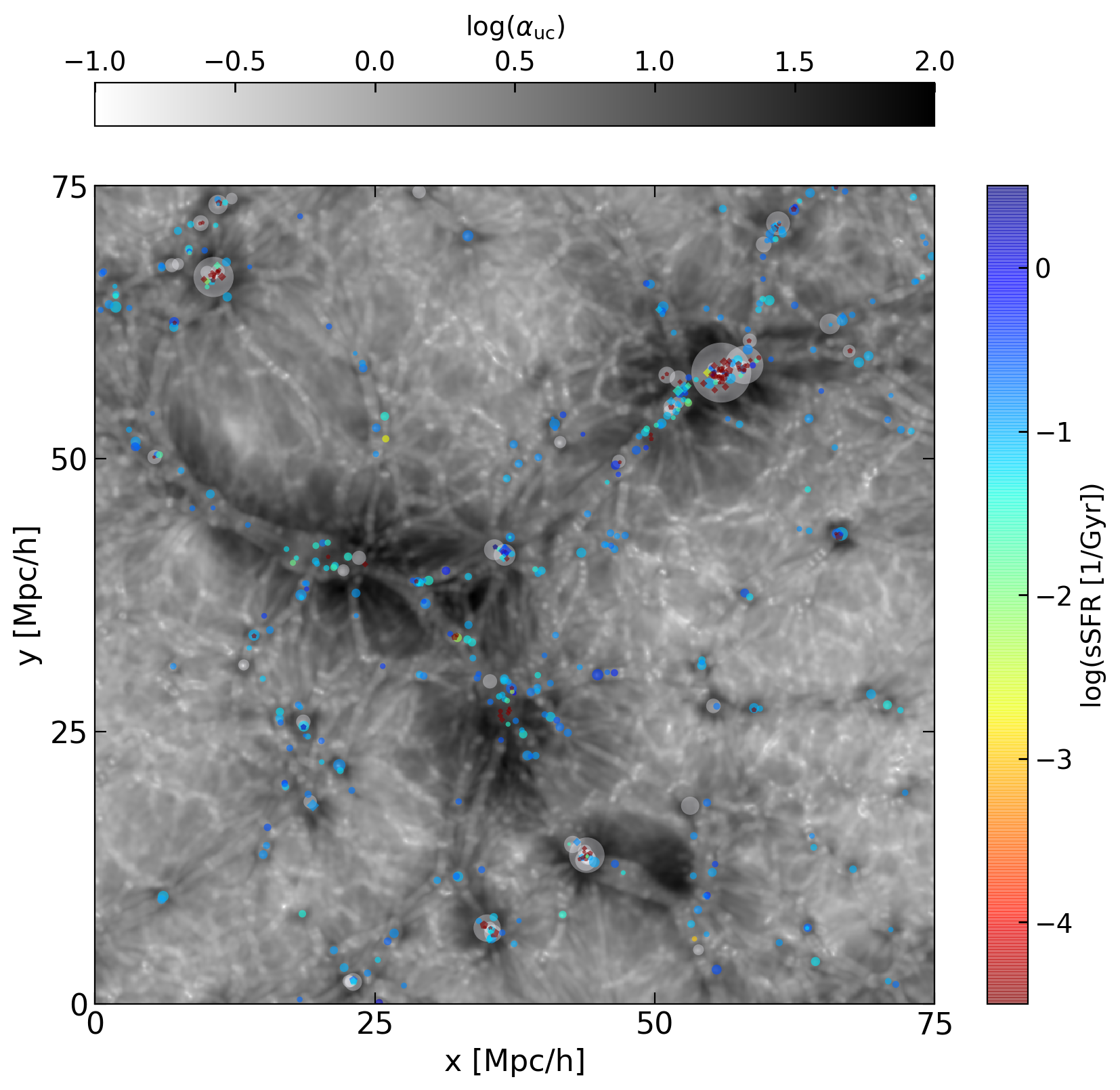}
\caption{Single-cell (`razor thin') slice through the TNG100 box, showing galaxies (circles, pentagons, and diamonds for centrals, satellites, and splashbacks, respectively) on a background showing overdensity $\delta$ \emph{(left panel)} and a modified tidal anisotropy $\alpha_{\rm uc}=\alpha\times(1+\delta)^{0.51}$ \emph{(right panel)} evaluated on a $1024^3$ grid with $250\kpc$ Gaussian smoothing. $\alpha_{\rm uc}$ is defined to have zero correlation with $\delta$ \citep[see][]{azpm19}. We display it to illustrate the difference between density and the tidal field at a fixed small smoothing scale; the variable $\alpha$ is thus a non-linear, scale-dependent combination of $\alpha_{\rm uc}$ and $\delta$.
Galaxies lie within $2\Mpc$ of the density slice with sizes scaled to $4R_{\rm vir}$ of the host (sub)halo (the adaptive scale at which $\alpha_{\rm peak}$ is extracted for each galaxy). 
For objects with $v_{\rm peak}\leq190\kms$, marker colours in the \emph{left (right) panel} show the host-centric $\alpha_{\rm peak}$ (galaxy sSFR); larger objects are transparent and white.}
\label{fig:visual}
\end{figure*}

Implementations of these ideas using empirical techniques lead to excellent descriptions of the dependence of galaxy clustering on sSFR and $M_\ast$ (equivalently, colour and luminosity) \citep{ss09,zehavi+11}. $M_\ast$-dependent clustering, in particular, is succinctly captured by subhalo abundance matching (SHAM) of $M_\ast$ and $v_{\rm peak}$, the largest value of the galaxy's maximum circular velocity over cosmic time \citep{rwtb13}, which is a mass proxy effectively set at the epoch of the last major merger \citep{behroozi+14}.

As regards the influence of the cosmic web, simulations indicate that cosmic filaments can channel cold gas into low-mass haloes \citep{kkwd05}. The local tidal environment also strongly influences halo mass accretion and mass loss in the vicinity of large objects, thus potentially regulating star formation \citep{hahn+09,behroozi+14}. The corresponding `assembly bias' imprints in galaxy clustering, however, remain observationally elusive \citep[e.g.,][]{azpm19}. Indeed, simplified models of galaxy evolution which treat centrals and satellites separately \citep[e.g.,][]{dfo12,birrer+14} provide reasonable descriptions of star formation trends with environment. A tidal influence on star formation, \emph{having assumed the central-satellite dichotomy}, is therefore weak at best. 

In this \emph{Letter}, we present evidence for an alternate viewpoint, in which star formation activity can be tied to a \emph{continuous measure} of the degree of environmental influence. Using the IllustrisTNG simulation, we show that the central-satellite dichotomy is, in fact, better represented as a continuum of \emph{increasing anisotropy} of the local tidal environment of a galaxy as we progress from centrals to satellites to splashback objects.\footnote{Splashbacks are current-epoch centrals that passed through a larger host in the past. We give a more detailed definition below.} Moreover, within each category, the sSFR of low-mass objects steadily decreases with increasing tidal anisotropy. Thus, the degree of tidal anisotropy acts as an excellent environmental proxy which blurs the distinction between traditional (star forming) centrals and (quenched) satellites. We describe the simulation and our analysis in section~\ref{sec:illustris}, present our results in section~\ref{sec:results} and conclude with our arguments in section~\ref{sec:conclude}. Throughout, we refer to objects with sSFR $>0.2 \ {\rm Gyr}^{-1} \, (<0.01 \ {\rm Gyr}^{-1})$ as `star forming' (`quenched') \citep[e.g.,][]{genel+15}.

\section{Simulation \& Analysis} 
\label{sec:illustris}

We use the outputs of the cosmological hydrodynamical simulation TNG100 from the galaxy formation simulation suite IllustrisTNG, focusing on results at $z = 0$. TNG100 evolves a $75\,h^{-1} {\rm Mpc}$ periodic box using the \citet{Planck15} cosmology and subgrid models for galaxy formation physics including, among others, recipes for star formation, stellar feedback in the form of galactic winds, black hole formation and growth, and AGN feedback. 

The subgrid models successfully reproduce the observed stellar mass function, the evolving SFR density and SFR-$M_\ast$ relation, and the low-$z$ $M_\ast$- and colour-dependent clustering \citep{fTNGSpringel18,fTNGPillepich18,fTNGNelson18,fTNGMarinacci18,fTNGNaiman18}. A key element for many of these successes is a new AGN feedback model in which black holes with masses $\gtrsim 2\times 10^{8}\Msun$, found in galaxies with $M_\ast \gtrsim 10^{10.5}\Msun$ ($v_{\rm peak}\gtrsim190\kms$, see below), transition from purely thermal feedback to driving kinetic winds which rapidly quench star formation 
\citep{Weinberger17,Weinberger18}.

\subsection{Galaxy populations}
We identify galaxies using the group finder \textsc{subfind} \citep{Springel01}. Applying a (conservative) $\geq 300$ stellar particles cut for galaxies with well-resolved sSFR at $z=0$ results in $17893$ satellite galaxies hosted in subhaloes and $19066$  `central' galaxies hosted at the potential minimum of (group) hosts. For the latter, we use \textsc{sublink} \citep{RodriguezGomez15Sublink} `baryonic' merger trees -- constructed using only stellar particles and star forming gas to identify objects, which allows for robust tracking of orphan galaxies -- to follow each main progenitor back in time until $z=1$ (lookback time $\simeq$ 8 Gyr). Objects which were identified, in any snapshot during $1\geq z > 0$, as satellites of hosts with centrals at least 3 times more massive than the pre-merger progenitor, are then labelled splashbacks,\footnote{We expect $z=1$ to be a sufficiently early epoch to identify passage through a group host. We have checked that minor variations in the mass ratio do not affect our results.} while the remaining objects are designated centrals. 

We use $v_{\rm peak}$ to account for the influence of a galaxy's host (sub)halo in establishing environmental trends (see above), allowing for a simultaneous discussion of all galaxy types, by-passing the complication of tidal effects on the dark matter content of substructure \citep{chaves-montero+16}. For each galaxy, we use the main progenitor branch of the merger trees to extract at each epoch the maximum value $v_{\rm max}$ of its rotation curve, whose maximum over time yields $v_{\rm peak}$. We also separately treat objects with host $R_{\rm vir} < 60.4\kpc$ comoving for which we do not have reliable estimates of the tidal environment (see below). Excluding these yields $17666$ centrals, $9468$ satellites, and $1300$ splashbacks  in our final sample. For each object, we use values of $M_\ast$ and instantaneous sSFR in our analysis.

\begin{figure*} 
\centering
\includegraphics[width=0.8\textwidth,trim= 6 5 8 6,clip]{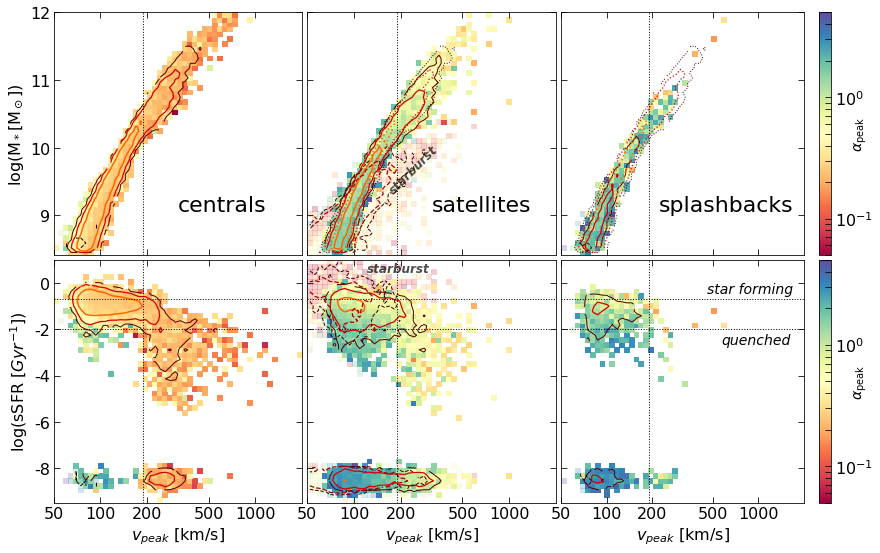}
\caption{Distribution of $v_{\rm peak}$ against $M_\ast$ \emph{(top panels)} and sSFR \emph{(bottom panels)} for central, satellite, and splashback galaxies identified in the Illustris TNG100 box (respectively, \emph{left, middle} and \emph{right panels}). Coloured histograms indicate the median value of peak tidal anisotropy $\alpha_{\rm peak}$, calculated for each object as in section~\ref{subsec:alpha}, overlayed with solid contours of equal galaxy number (orange, red and brown for 100, 20 and 5 objects, respectively). Dotted contours in the \emph{top middle} and \emph{top right panels} repeat the $M_\ast$-$v_{\rm peak}$ contours for central galaxies from the \emph{top left panel}. Dashed contours and transparent histograms indicate objects with $R_{\rm vir} < 60.4\kpc$ (see text). Clouds near $\log({\rm sSFR} / {\rm Gyr}^{-1})=-8.5$ in the \emph{bottom panels} represent objects with sSFR = 0 in the simulation. Vertical dotted line in each panel indicates $v_{\rm peak}=190\kms$ which we use to segregate small and large objects. Horizontal dotted lines in the \emph{bottom panels} indicate the thresholds used to define actively star forming and quenched galaxies.}
\label{fig:main}
\end{figure*}

\subsection{Local tidal environment}
\label{subsec:alpha}

For a galaxy in a host with virial radius $R_{\rm vir}$, we first characterise its local environment using the \emph{tidal anisotropy} scalar $\alpha$ extracted from the host-centric tidal tensor Gaussian-smoothed at scale $R_{\rm h}\equiv4R_{\rm vir}/\sqrt{5}$ \citep{phs18a}. In terms of the eigenvalues $\lambda_1$, $\lambda_2$, $\lambda_3$ of the smoothed tidal tensor $\psi_{,ij}=\p^2\psi/\p x^i\p x^j$ (where the normalised gravitational potential $\psi$ satisfies the Poisson equation $\nabla^2\psi=\delta$), $\alpha$ is defined as
\be
\alpha\equiv\sqrt{q^2} / (1+\delta)\,,
\label{eq:alpha-def}
\ee

where $\delta=\lambda_1+\lambda_2+\lambda_3$ is the total matter overdensity and $q^2=(1/2)\left[(\lambda_1-\lambda_2)^2+(\lambda_2-\lambda_3)^2+(\lambda_3-\lambda_1)^2\right]$ is the tidal shear, so that $\alpha\geq0$. Haloes in filaments have $\alpha\geq0.5$ while subhaloes span both larger and smaller values depending on their host-centric distance \citep{paranjape20}. 

To determine $\alpha$ for each galaxy, we first calculate the overdensity field $\delta(\bvec{x})$ in the full TNG100 volume on a $1024^3$ grid using cloud-in-cell interpolation, then Fourier transform and Gaussian-smooth this on 17 (approximately logarithmically spaced) smoothing scales $R_{\rm G}$ between $108\kpc$ and $5\Mpc$,\footnote{The smallest scale is set by the grid size and in turn sets the minimum host $R_{\rm vir}$ we can use.} to obtain the $k$-space fields $\delta(\bvec{k};R_{\rm G})=\delta(\bvec{k})\e{-k^2R_{\rm G}^2/2}$. The smoothed tidal tensor fields $\psi_{,ij}(\bvec{x};R_{\rm G})$ are obtained as the inverse Fourier transform of $(k_ik_j/k^2)\delta(\bvec{k};R_{\rm G})$ for each $R_{\rm G}$. These are interpolated spatially and across smoothing scales to obtain the (sub)halo-centric $\psi_{,ij}(\bvec{x}_{\rm  h};R_{\rm h})$ at the location $\bvec{x}_{\rm h}$ of the galaxy's host at scale $R_{\rm h}$ (see above), diagonalising which finally leads to a value of $\alpha$ for the galaxy using \eqn{eq:alpha-def}. 

\citet{rphs19} showed that this definition of $\alpha$ with its choice of adaptive smoothing scale is an excellent indicator of halo assembly bias, particularly for objects whose current tidal environment is the most extreme they have experienced. This motivates us to focus on $\alpha_{\rm peak}$, the maximum of $\alpha$ for an object over cosmic time, as a proxy for environmental influence. We calculate the tidal tensor and (sub)halo-centric $\alpha$ as discussed above in each snapshot between $1\geq z>0$ and extract $\alpha_{\rm peak}$ for each object (similarly to $v_{\rm peak}$) along with the redshift $z_{\alpha{\rm peak}}$ at which the maximum occurred.

Figure~\ref{fig:visual} visualises a single-cell slice through the simulation, chosen to contain a few clusters and some filaments, with galaxies shown by the markers on a background of the density $\delta$ and a modified tidal anisotropy $\alpha_{\rm uc}$ in the \emph{left} and \emph{right panels}, respectively (see the figure caption for details).

\section{Results} 
\label{sec:results}
\noindent

Figure~\ref{fig:main} shows the distribution of $v_{\rm peak}$ against $M_\ast$ (sSFR) in the \emph{top (bottom) panels} for centrals, satellites and splashbacks, with coloured histograms showing the median value of $\alpha_{\rm peak}$ in each bin, overlayed with solid contours of equal number. For each galaxy type, in the \emph{top panels} we see a tight correlation between $M_\ast$ and $v_{\rm peak}$. The dotted contours in the \emph{top middle} and \emph{right panels} repeat the $M_\ast$-$v_{\rm peak}$ relation for centrals; this is evidently identical to the relation for satellites and splashbacks.\footnote{Dashed contours and transparent histograms (prominent in the middle panels) indicate objects with $R_{\rm vir} < 60.4\kpc$ comoving, for which we can only measure $\alpha$, and hence $\alpha_{\rm peak}$, at the grid scale $R_{\rm G}=108\kpc$ comoving. Of these, the low-$\alpha_{\rm peak}$, high-sSFR satellites (labelled `starburst') scatter the most off the $M_*$-$v_{\rm peak}$ relation. These are possibly recent major mergers, which could explain why \citet{chaves-montero+16} found that defining $v_{\rm peak}$ while an object is dynamically relaxed yields the tightest correlation with $M_\ast$.} The most striking feature of the \emph{top panels}, however, is the steady increase in typical $\alpha_{\rm peak}$ values as we progress from centrals $\to$ satellites $\to$ splashbacks.

The \emph{bottom panels} of Figure~\ref{fig:main} show the bimodality of sSFR as a function of $v_{\rm peak}$. Objects with sSFR $=0$ in the simulation are visualised as narrow clouds near the bottom of each panel.\footnote{We artificially assigned a small value of sSFR to these objects by adding a Gaussian scatter of 0.2 dex around $\log({\rm sSFR} / {\rm Gyr}^{-1})=-8.5$.} We see the usual trend of star forming (quenched) objects being mostly centrals (satellites). Splashbacks form a small, predominantly quenched population. Additionally, the sSFR-$v_{\rm peak}$ distribution coloured by $\alpha_{\rm peak}$ shows some remarkable features. 

At $v_{\rm peak}\gtrsim190\kms$ (or $M_\ast\gtrsim10^{10.5}\Msun$ as inferred from the \emph{top panels}), most objects are in isotropic environments ($\alpha_{\rm peak}\lesssim0.3$ for centrals, somewhat larger for satellites). The downturn and suppression of sSFR apparent for centrals and satellites is due to AGN activity in massive haloes (see section~\ref{sec:illustris}), uncorrelated with tidal environment. 

At $v_{\rm peak}\lesssim190\kms$, on the other hand, AGN do not play a role in the IllustrisTNG model. Instead, the peak tidal anisotropy $\alpha_{\rm peak}$ takes centre-stage:
\begin{enumerate}
\item Star forming centrals and satellites \emph{exclusively occupy} low-$\alpha_{\rm peak}$ environments. 
\item As we decrease the sSFR of centrals and satellites at fixed $v_{\rm peak}$, $\alpha_{\rm peak}$ steadily increases, ending in objects with sSFR $=0$ which predominantly live in high-$\alpha_{\rm peak}$ environments. Focusing on \emph{non star forming} objects (i.e. ${\rm sSFR} < 0.2 \ {\rm Gyr}^{-1}$), when sSFR $\neq0$ we find Spearman rank correlation coefficients between $\alpha_{\rm peak}\leftrightarrow$ sSFR of $\sim-0.1\,(-0.35)$ for centrals (satellites), independent of $v_{\rm peak}$. For objects with sSFR $=0$, the median $\alpha_{\rm peak}$ values are $1.75 \,(2.45)$ for centrals (satellites).
\item Although splashbacks are predominantly quenched and in overall higher-$\alpha_{\rm peak}$ environments, the same trends are apparent. The $\alpha_{\rm peak}\leftrightarrow$ sSFR Spearman coefficient for \emph{non star forming} splashbacks with sSFR $\neq0$ is $\sim-0.6$ and the median $\alpha_{\rm peak}$ for sSFR $=0$ objects is $2.9$. In fact, we find that the small population of sSFR $=0$ \emph{centrals} is similar to the sSFR $=0$ splashbacks as regards the distribution of $\alpha_{\rm peak}$ and hence location in the cosmic web. These objects are possibly splashbacks misclassified as centrals due to resolution limitations of the merger tree. Thus, the trend of overall $\alpha_{\rm peak}$ increasing from centrals to splashbacks comes back full circle for the quenched population.
\end{enumerate}

Figure~\ref{fig:visual} reflects the trends discussed above, particularly the fact that the lowest star formation is restricted to objects with the highest anisotropy.  We have checked that similar trends are \emph{not} apparent when using $\delta$ smoothed at the same adaptive scale as $\alpha$ (and hence $\alpha_{\rm peak}$), but do emerge for $\delta$ at much larger ($\sim4\times$) smoothing scales, consistent with previous results on halo assembly bias \citep{phs18a}.

Figure~\ref{fig:qfVsalphapk} shows that the quenched fraction $f_{\rm q}$ of galaxies of each type (a) increases steeply with $\alpha_{\rm peak}$ and (b) has similar magnitude for all types if we exclude objects with $z_{\alpha{\rm peak}}<0.15$ for which $\alpha$ achieved its maximum within the last 2 Gyr. We discuss these trends below.

\section{Discussion} 
\label{sec:conclude}

The steady increase of local peak tidal anisotropy $\alpha_{\rm peak}$ from centrals to satellites and from star forming to quenched objects in Figure~\ref{fig:main} is suggestive of a continuous hierarchy of tidal environments that spans the low-mass central-satellite dichotomy. This idea is reinforced by the fact that splashback objects (which spatially mimic centrals but are physically closer to satellites) emerge as an extreme population with high $\alpha_{\rm peak}$ and low sSFR. This is also consistent with previous results showing that the tidal influence of a satellite's group host extends well outside its $R_{\rm vir}$ \citep{behroozi+14,diemer20a,blp20}. Since the local tidal anisotropy is also known to have a strong positive correlation with large-scale clustering \citep{phs18a}, \emph{the continuum of tidal anisotropy naturally accounts for the environment-dependence of the sSFR-$M_\ast$ bimodality}.

\begin{figure} 
\centering
\includegraphics[width=0.4\textwidth,trim= 0 20 38 60,clip]{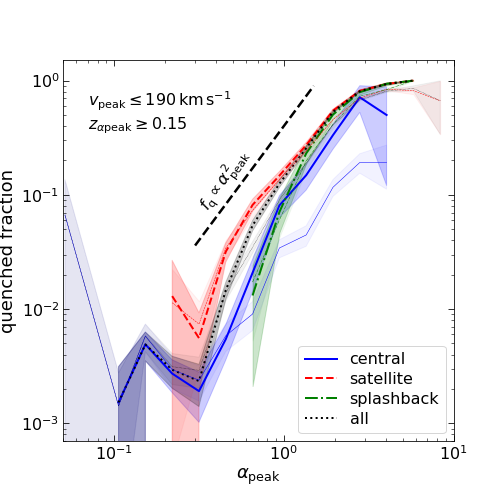}
\caption{Quenched fraction $f_{\rm q}$ as function of $\alpha_{\rm peak}$ for objects with $v_{\rm peak}\leq190\kms$. For each galaxy type, we bin objects by $\alpha_{\rm peak}$ and evaluate $f_{\rm q}$ as the ratio of number of objects with sSFR $<0.01 \ {\rm Gyr^{-1}}$ to the total number in each $\alpha_{\rm peak}$ bin. Thin lines are for the full sample while thick lines show results when excluding objects with $z_{\alpha{\rm peak}}<0.15$ (lookback time $\lesssim2$ Gyr). Bands show errors from 200 bootstrap samples. Black dashed line segment illustrates the behaviour $f_{\rm q}\propto\alpha_{\rm peak}^2$ for comparison.}
\label{fig:qfVsalphapk}
\end{figure}

One might also ask whether $\alpha_{\rm peak}$ can explain the bimodality itself. If so, the quenched fraction $f_{\rm q}$ of each galaxy type should be set by $\alpha_{\rm peak}$ alone, with galaxies being preferentially quenched in anisotropic environments, regardless of their identity as central, satellite, or splashback.  Figure~\ref{fig:qfVsalphapk} shows that $\alpha_{\rm peak}$ is, in fact, a strong indicator of $f_{\rm q}$ for each type. When considering the full sample (thin lines), however, $f_{\rm q}$ for centrals is about a factor $5$ lower than that of satellites and splashbacks. Such a deficit is not surprising, since galaxy quenching is expected to occur over timescales of $\lesssim2$ Gyr after the event that set $\alpha_{\rm peak}$ \citep[such as accretion onto a cluster; see, e.g.,][]{mzhs19}. We can allow for a delay of this magnitude by excluding objects for which $z_{\alpha\rm{peak}}\lesssim0.15$. The thick lines in the figure show that \emph{applying such a cut indeed removes most of the differences between $f_{\rm q}$ for the different galaxy types,} with the effect being strongest for the centrals. The excluded centrals would be predicted to quench their star formation within $\sim2$ Gyr. Indeed, we find that the median gas fraction of centrals with $z_{\alpha{\rm peak}}\leq0.15$ and $\alpha_{\rm peak}>1$ is $\sim25\%$ smaller than of all centrals with $v_{\rm peak}\leq190\kms$, indicating the onset of gas stripping.

All of this strongly suggests that, just as $v_{\rm peak}$ is an excellent indicator of stellar mass regardless of galaxy type, the relevant indicator of star formation is $\alpha_{\rm  peak}$. The combination of $v_{\rm peak}$ and $\alpha_{\rm peak}$ essentially erases the dichotomy between centrals and satellites as regards star formation activity: star forming centrals on one hand and quenched satellites on the other are simply extremes in a continuum of tidal environments. SHAM with $(v_{\rm peak},\alpha_{\rm peak})$ will therefore be very interesting to pursue. 

We emphasize that the link between sSFR and $\alpha_{\rm peak}$ is a statistical one; the physics of star formation inherently occurs at scales much smaller than $4R_{\rm vir}$ and is likely only indirectly affected by $\alpha_{\rm peak}$. We also caution that some of our results (such as the specific value $v_{\rm peak}\simeq190\kms$ for the onset of AGN effects) could represent subgrid choices in the IllustrisTNG galaxy formation model. We expect, however, that our overall conclusions are robust to subgrid modelling, since they rely on broad properties of galaxies and the cosmic web. 

The shift from a dichotomy to a continuum potentially affects the understanding of halo models as well as semi-analytical and empirical models of galaxy evolution. For example, our results open up the exciting possibility of using the $\alpha_{\rm peak}\leftrightarrow$ sSFR connection (calibrated in small-volume galaxy formation simulations) to easily model the secondary clustering bias caused by an sSFR-dependent galaxy sample selection. More direct observational tests would require overcoming the challenges of stochastic biasing and non-linear redshift space effects when estimating the local tidal environment of galaxies.
We will explore these ideas in future work.

\vskip 0.1in
\begin{spacing}{0.8}
\noindent
\textbfit{Acknowledgments:} 
{\small
We thank R. Srianand for useful conversations, R. Angulo for comments on an earlier draft, and V. Rodriguez-Gomez for kindly providing us the baryonic version of the \textsc{sublink} merger trees.
JZ and OH acknowledge funding from the European Research Council (ERC) under the European Union's Horizon 2020 research and innovation programme, Grant agreement No. 679145 (COSMO-SIMS).
AP acknowledges funding from the Associateship Scheme of ICTP, Trieste and the Ramanujan Fellowship of the DST, Govt of India.
}
\end{spacing}

\vskip 0.1in
\begin{spacing}{0.8}
\noindent
\textbfit{Data availability:}
{\small
IllustrisTNG data is publicly available at \href{https://tng-project.org}{tng-project.org} \citep{Nelson19TNGDR}. 
}
\end{spacing}

\bibliography{masterRef,citations}

\label{lastpage}

\end{document}